\documentclass[12pt,a4paper]{article}

\usepackage[left=2.5cm,top=2.5cm,bottom=2.5cm,right=2.5cm]{geometry}
\usepackage{graphicx}
\usepackage{amssymb}
\usepackage{amsthm}
\usepackage{amsmath}
\usepackage{amsfonts}
\usepackage{algorithm, algorithmic}
\usepackage{color}

\usepackage{cite}

\usepackage{enumitem}

\newtheorem{thm}{Theorem}[section]

\newtheorem{lem}[thm]{Lemma}

\theoremstyle{definition}

\theoremstyle{remark}
\newtheorem{rem}{Remark}[section]

\numberwithin{equation}{section}

\newcommand{\lemref}[1]{Lemma~\ref{#1}}

\DeclareMathOperator{\idiv}{\,{\bf div } \,}
\DeclareMathOperator{\imod}{\,{\bf mod } \,}

\begin{document}

\title{Sorting distinct integer keys using in-place associative sort}


\author{A. Emre CETIN \\
email: aemre.cetin@gmail.com}

\maketitle

\begin{abstract}

{\em In-place associative integer sorting} technique was proposed for integer lists which requires only constant amount of additional memory replacing bucket sort, distribution counting sort and address calculation sort family of algorithms. The technique was explained by the analogy with the three main stages in the formation and retrieval of memory in cognitive neuroscience which are (i) {\em practicing}, (ii) {\em storing} and (iii) {\em retrieval}. 

In this study, the technique is specialized with two variants one for read-only integer keys and the other for modifiable integers. Hence, a novel algorithm is obtained that does not require additional memory other than a constant amount and sorts faster than {\em all} no matter how large is the list provided that $m = \mathcal{O} (n \log n)$ where $m$ is the range and $n$ is the number of keys (or integers).

\end{abstract}


\section{Introduction}\label{sec:intro}


The main difficulties of all distributive sorting algorithms is that, when the keys are distributed using a hash function according to their content, several keys may be clustered around a loci, and several may be mapped to the same location. These problems are solved by inherent three basic steps of associative sort~\cite{ecetin} (i) {\em practicing}, (ii) {\em storing} and (iii) {\em retrieval} which are the three main stages in the formation and retrieval of memory in cognitive neuroscience. The technique assumes that associations are between the integers in the list space and the nodes in an imaginary linear subspace that spans a predefined interval of range of integers. The imaginary subspace can be defined anywhere on the list space $S[0\ldots n-1]$ provided that its boundaries do not cross over that of the list. The range of the interval that the imaginary subspace spans is upper bounded with the number of integers $n$ but may be smaller and can be located anywhere making the technique in-place, i.e., beside the input list, only a constant amount of memory locations are used for storing counters and indices. Furthermore, this definition reveals the asymptotic power of the technique with increasing $n$ with respect to the range of integers, as well. 

An association between a integer and the imaginary subspace is created by a node using a monotone bijective hash function that maps the integers in the predefined interval to the imaginary subspace. The process of creating a node by mapping a distinct integer to the imaginary subspace is ``practicing a distinct integer of an interval''. Since imaginary subspace is defined on the list space, this is just swapping. Once a node is created, the redundancy due to the association between the integer and the position of the node releases the word allocated for the integer in the physical memory except one bit which tags the word as a node for interrogation. All the bits of the node except the tag bit can be cleared and used to encode any information. Hence, they are the ``record'' of the node and the information encoded into a record is the ``cue'' by which cognitive neuro-scientists try to describe how the brain recalls the next item in the order during retrieval. For instance, it will be foreknown from the tag bit that a node has already been created while another occurrence of that particular integer is being practiced giving the opportunity to count other occurrences. The process of counting other occurrences of a particular integer is ``practicing all the integers of an interval'', i.e., rehearsing used by cognitive neuro-scientists to describe how the brain manipulates the sequence before storing in short (or long) term memory. On the other hand, the tag bit discriminates the word as node and the position of the node lets the integer be retrieved back from the imaginary subspace using the inverse hash function.  

Practicing does not need to alter the value of other occurrences, i.e., only the first occurrence is altered while being practiced from where a node is created. All other occurrences of that particular integer remain in the list space but become meaningless. Hence they are ``idle integers''. On the other hand, practicing does not need to alter the position of idle integers as well, unless another distinct integer creates a node exactly at the position of an idle integer while being practiced. In such a case, the idle integer is moved to the former position of the integer that creates the new node. This makes associative sort unstable, i.e., equal integers may not retain their original relative order. 

Once all the integers in the predefined interval are practiced, the nodes that are dispersed in the imaginary subspace with relative order are clustered in a systematic way, i.e., the distance between the nodes are closed to a direction retaining their relative order. This is the {\em storing} phase of associative sort where the received, processed and combined information to construct the sorted permutation of the practiced interval is stored in the short-term memory. When the nodes are moved towards a direction, it is not possible to retain the association between the imaginary subspace and list space. However, the record of a node can be further used to encode the absolute position of that node as well, or maybe the relative position or how much that node is moved relative to its absolute or relative position during storing. Unfortunately, this requires that a record is enough to store both the position of the node and the number of idle integers practiced by that node. However, as explained earlier, further associations can be created using the idle integers that were already practiced by manipulating either their position or value or both. Hence, if the record is enough, it can store both the positional information and the number of idle integers. If not, an idle integer can be associated accompanying the node to supply additional space for it for the positional information.

Finally, the sorted permutation of the practiced interval is constructed in the list space, using the stored information in the short-term memory. This is the {\em retrieval} phase of associative sort that depends on the information encoded into the record of a node. If the record is enough, it stores both the position of the node and the number of idle integers. If not, an associated idle integer accompanying the node stores the position of the node while the record holds the number of idle integers. The positional information cues the recall of the integer using the inverse hash function. This is ``integer retrieval'' from imaginary subpace. Hence, the retrieved integer can be copied on the list space as much as it occurrs.

Hence, moving through nodes that represent the start and end of practiced integers as well as retaining their relative associations with each other even when their positions are altered by cuing allow the order of integers to be constructed in linear time in-place.

From complexity point of view, associative sort shows similar characteristics with bucket sort~\cite{mahmoud:2000,Cormen} and distribution counting sort~\cite{Seward,Feurzig}. It sorts $n$ integers $S[0\ldots n-1]$ each in the range $[0,m-1]$ using $\mathcal{O}(1)$ extra space in $\mathcal{O}(n+m)$ time for the worst, $\mathcal{O}(m)$ time for the average (uniformly distributed integers) and $\mathcal{O}(n)$ time for the best case. The ratio $\frac{m}{n}$ defines the efficiency (time-space trade-offs) of the algorithm letting very larges lists to be sorted in-place. 

\subsection{Specialized Version for Distinct Integers}

If it is known that all the integers of the list are distinct, associative sorting technique can be specialized because there is only one integer that can be practiced and mapped to a location creating a node. This means that 2 solutions are possible for lists of distinct keys. The first one is for read-only keys and instead of tagging the word as node using its most significant bit (MSB), the key itself can be used to tag the word ``implicitly'' as node without modifying it, since when a key is mapped to the imaginary subspace, it will always satisfy the monotone bijective hash function. The keys are ``implicitly practiced'' in this case. Hence, storing phase is enough to obtain the sorted permutation of the practiced interval cancelling the retrieval phase. In each iteration only the keys that fall into the range $[\delta, \delta+n-1]$ can be sorted where $\delta$ is the minimum of the list of that iteration. It should be noted that, this variant is suitable for sorting a list $S$ of $n$ {\em elements}, $S[0 \ldots n-1]$ each have an integer {\em key} where the problem is to sort the elements of the list according to their integer keys. 

The other scenario is that, when a distinct integer is mapped to the imaginary subspace, its record can be used to improve the interval of range of integers that are practiced. During storing, each node is clustered at the beginning of the list together with its record retaining its relative order with respect to others. At this point, we need $\log n$ bits of the record to encode the node's absolute position to cue the retrieval of the integer from the imaginary subspace. But the tag bit can be released during storing phase since we only need how many nodes are stored at the beginning of the list in total. Hence we can use for instance the least significant $w-\log n$ bits of a record during practicing for any other purpose. It is immediate from this definition that a monotone bijective super hash function can be used during practicing. It should be noted that, this variant is suitable for sorting a {\em list} $S$ of $n$ {\em integers}, $S[0 \ldots n-1]$ where the problem is to sort the integers in ascending or descending order. 

With this introductory information, the contributions of this study are,
\begin{description}[leftmargin=0pt]
\item[{\bf A practical algorithm}] that sorts a list of $n$ elements $S[0\ldots n-1]$ each have a {\em read-only} {\em distinct} integer key in the range $[0,m-1]$ using $\mathcal{O}(1)$ extra space in $\mathcal{O}(n+m)$ time for the worst, $\mathcal{O}(m)$ time for the average (uniformly distributed keys) and $\mathcal{O}(n)$ time for the best case. Therefore, the ratio $\frac{m}{n}$ defines the efficiency (time-space trade-offs) letting very large lists to be sorted in-place. 

Practical comparisons with $\Omega(n \log n)$ quick sort\cite{Hoare, sedgewick:algorithms_in_C} and merge sort\cite{Anonymous_1} and heap sort\cite{Williams,Levitin} which take $\mathcal{O}(n \log n)$ time on all inputs showed that associative sort is superior than all (up to 20 times) provided that $\frac{m}{n} \le c\log n$ where $c \approx 4$ for both heap sort and merge sort. Quick sort gave worser results ($c \approx 8$) for distinct keys. These results are consistent with $m$ calculated theoretically making average case time complexity of the algorithm less than lower-bound of comparison-based sorting algorithms, i.e., $\mathcal{O}(m) < \Omega (n \log n)$. Another very important meaning of this inequality is that, since it does not require additional memory space other than a constant amount, no matter how large is the list, the proposed algorithm will sort faster than {\em all} provided that $m = \mathcal{O} (n \log n)$.

Practical comparisons with instable distribution counting sort (which shows better performance than stable one) showed that associative sort is superior in every case. This is expectable considering memory allocation overload of distribution counting sort since time-complexities of both algorithms are same. The performance of the algorithm is even {\em asymptotically} better than 2 lines of code referred in textbooks for sorting $n$ distinct integers from $[0\ldots n - 1]$ with indexing an auxiliary list $B$ of the same size as the input $A$ with keys of $A$ by $B[A[i]]=A[i]$ for $i=0,1,\ldots,n-1$, and then reconstructing sorted permutation of $A$ back by $A[i]=B[i]$ for $i=0,1,\ldots,n-1$, which is due to time-consuming memory allocation of the auxiliary list. 

Associative sort for read-only distinct keys has been compared with radix sort~\cite{knuth:vol3,mahmoud:2000,sedgewick:algorithms_in_C, Cormen} and bucket sort, as well. The results showed that it is superior than radix sort when $\frac{m}{n} \le 8$ and 2 times faster than bucket sort for $n$ {\em distinct} integer keys $S[0...n-1]$ each in the range $[0, n-1]$. 

Finally, the dependency of the efficiency of associative sort on the distribution of the keys is only $\mathcal{O}(n)$ which means it replaces all the methods based on address calculation~\cite{Isaac,Tarter,Flores,Jones,Gupta,Suraweera}, that are known to be very efficient when the keys have known (usually uniform) distribution and require additional space more or less proportional to $n$~\cite{knuth:vol3}. 

\item[{\bf A practical algorithm}] that sorts $n$ {\em modifiable} {\em distinct} integers $S[0\ldots n-1]$ each in the range $[0,m-1]$ using $\mathcal{O}(1)$ extra space with an efficiency improvement of $\frac{m}{n(w-\log n)}$. In other words, if $\frac{m}{n} \le {w-\log n}$ the complexity of the algorithm is strictly $\mathcal{O}(n)$. Otherwise, it sorts the integers using $\mathcal{O}(1)$ extra space in $\mathcal{O}(n+\frac{m}{w-\log n})$ time for the worst, $\mathcal{O}(\frac{m}{w-\log n})$ time for the average (uniformly distributed integers) and $\mathcal{O}(n)$ time for the best case. Similarly, the efficiency can be represented with $\frac{m}{n} \le c(w-\log n)$ where the constant $c > 1$ is determined by the other sorting algorithms. When modifiable version is compared with read-only version, it has been observed that up to $\frac{m}{n} \le 10$ read-only version is more efficient possibly due to bitwise operations involved in the modifiable version. Afterwards modifiable version becomes more efficient than read-only version augmenting it.

\end{description}

Finally, the technique requires at least $0$, at most $2n-k$ swaps, where $k>0$ is the number of iteration (or dept of recursion) to complete the sorting which sets the lower bound for number of data movements to complete a sorting.

Even omitting its space efficiency for a moment, associative sort asymptotically outperforms all content based sorting algorithms when $n$ is large relative to $m$.


\section{Definitions}\label{sec:pre}
The definition of {\em integer key sorting} is: given a {\em list} $S$ of $n$ {\em elements}, $S[0 \ldots n-1]$ each have an integer {\em key}, the problem is to sort the elements of the list according to their integer keys. To prevent repeating statements like ``integer of the element $S[i]$'', $S[i]$ is used to refer the integer. 

The definition of {\em integer sorting} is: given a {\em list} $S$ of $n$ {\em integers}, $S[0 \ldots n-1]$, the problem is to sort the integers in ascending or descending order.

The notations used throughout the study are: 
\begin{enumerate} [label=({\roman{*}}), nosep]
\item Universe of integers is assumed $\mathbb{U} = [ 0 \ldots 2^{w}-1]$ where $w$ is the fixed word length.

\item Maximum and minimum integers of a list are, $\max (S) = \max(a \vert a \in S)$ and $\min (S) = \min(a \vert a \in S)$, respectively. Hence, range of the integers is, $m = \max (S) - \min (S) + 1$.

\item The notation $B \subset A$ is used to indicated that $B$ is a proper subset of $A$.

\item For two lists $S_{1}$ and $S_{2}$, $\max (S_{1}) < \min (S_{2})$ implies $S_{1} < S_{2}$.

\end{enumerate}

\begin{description}[leftmargin = 0pt]

\item[{\bf Universe of Integers.}] When an integer is first practiced, a node is created releasing $w$ bits of the integer free. One bit is used to tag the word as a node. Hence, it is reasonable to doubt that the tag bit limits the universe of integers because all the integers should be untagged and in the range $[0,2^{w-1}-1]$ before being practiced. But, we can,
\begin{enumerate}[label=(\roman{*}), itemindent = * , nosep]
\item partition $S$ into $2$ disjoint sublists $S_1 < 2^{w-1} \le S_2$ in $\mathcal{O}(n)$ time with well known in-place partitioning algorithms as well as stably with~\cite{Katajainen},
\item shift all the integers of $S_2$ by $-2^{w-1}$, sort $S_1$ and $S_2$ associatively and shift $S_2$ by $2^{w-1}$.
\end{enumerate}
There are other methods to overcome this problem. For instance, 
\begin{enumerate}[label=(\roman{*}), itemindent = * , nosep]
\item sort the sublist $S[0\ldots (n/ \log n)-1]$ using the optimal in-place merge sort~\cite{Salowe},
\item compress $S[0\ldots (n/ \log n)-1]$ by Lemma~1 of~\cite{Franceschini_1} generating $\Omega(n)$ free bits,
\item sort $S[(n/ \log n)\ldots n-1]$ associatively using $\Omega(n)$ free bits as tag bits,
\item uncompress $S[0\ldots (n/ \log n)-1]$ and merge the two sorted sublists in-place in linear time by~\cite{Salowe}.
\end{enumerate}

\item [{\bf Number of Integers.}] If practicing a distinct integer lets us to use $w-1$ bits to practice other occurrences of that integer, we have $w-1$ free bits by which we can count up to $2^{w-1}$ occurrences including the first integer that created the node. Hence, it is reasonable to doubt again that there is another restriction on the size of the lists, i.e., $n \le 2^{w-1}$. But a list can be divided into two parts in $\mathcal{O}(1)$ time and those parts can be merged in-place in linear time by~\cite{Salowe} after sorted associatively.

\end{description}

It should be noted that these restrictions are only valid for the variant proposed for modifiable integers. Hence, for the sake of simplicity, it will be assumed that $n \le 2^{w-1}$ and all the integers are in the range $[0,2^{w-1}-1]$ throughout the study.


\section{Basics of Associative Sort} \label{sec:basics}

Given $n$ {\em distinct} integers $S[0\ldots n-1]$ each in the range $[u, v]$, if $m=n$, the sorted permutation of the list can be represented with two parameters ($2 \log \mathbb{U}$ bits) one of which is the initial address of the sequential memory separated for the list (accessed by $S[0]$) in the RAM and the other is the $\delta=u$. The $i$th integer of the sorted list can be calculated by $S[i]=i+\delta$ and the actual value at $i$th location is meaningless for this calculation. Hence, if $S$ is the sorted permutation, then there is a bijective relation between each integer and its position, i.e., $i=S[i]-\delta$. From contradiction, if $S$ is not the sorted permutation, $i \ne S[i]-\delta$ implies that the integer $S[i]$ is not at its exact location. Its exact location can be calculated with $j=S[i]-\delta$. Therefore, this monotone injective hash function that maps the integers to $j \in [0,n-1]$ can sort the list in $\mathcal{O}(n)$ time using $\mathcal{O}(1)$ constant space. This is cycle leader permutation where $S$ is re-arranged by following the cycles of a permutation $\pi$. First $S[0]$ is sent to its final position $\pi(0)$ (calculated by $j=S[i]-\delta$). Then the element that was in $\pi(0)$ is sent to its final position $\pi(\pi(0))$. The process proceeds in this way until the cycle is closed, that is until the integer to position $0$ is found which means the association $0 = S[0] - \delta$ is constructed between the first integer and its position. Then the iterator is increased to continue with the integer of $S[1]$. At the end, when all the cycles of $S[i]$ for $i=0,1..,n-1$ are processed, all the integers are moved to their exact position and the association $i = S[i] - \delta$ is constructed between the integers and their positions, i.e., the sorted permutation of the list is obtained. 

If we look at this approach closer, we can interpret the technique entirely different. That is, we are indeed creating an imaginary subspace $Im[0\ldots n-1]$ over $S[0\ldots n-1]$ where the relative basis of this imaginary subspace coincides with that of the list space in the physical memory. The imaginary subspace spans a predefined interval of the range of integers depending on $n$. Since $m=n$, it spans the entire range of integers. The association between the list space and the imaginary subspace is created by a node using the monotone bijective hash function $i = S[i] - \delta$ that maps a particular integer to the imaginary subspace. When a node is created for a particular integer, the redundancy due to the association between the integer and the position of the node releases the word allocated for the integer in the physical memory. Hence, we can clear the node ($S[i]=0)$ and set its tag bit, for instance its most significant bit (MSB) to discriminate it as a node, and use the remaining $w-1$ bits of the node for any other purpose. When we want the integer back to list space from imaginary subspace, we can use the inverse of hash function and get the integer back by $S[i]=i + \delta$ to the list space. However, we don't use free bits of a node for other purposes in this case because it is known that all the integers are distinct and hence only one integer will be practiced at a location creating a node. Therefore, instead of tagging the word as node using its MSB, we use the integer itself to tag the word ``implicitly'' as node, since if a integer is mapped to the imaginary subspace, then it will always satisfy the monotone bijective hash function $i = S[i] - \delta$. Hence, the integers are ``implicitly practiced'' in this case.

\subsection{Sorting $n$ Distinct Read-Only Integer Keys} \label{sec:condition2}

The above definition immediately lets us to state that,
\begin{lem}\label{lem:sorting_seq_of_distinct}
Given a list $S$ of $n$ elements $S[0...n-1]$ each have a distinct integer {\em key} in the range $[u, v]$, $n_d$ elements of the list that have keys in the range $[\delta,\delta+n-1]$ with $\delta=\min(S)$ can be sorted associatively at the beginning of the list in $\mathcal{O}(n)$ time with at most $2n_d-1$ at least $0$ swaps, respectively using only $\mathcal{O}(1)$ constant space.
\end{lem}

\begin{proof}
Given $n$ distinct integer keys $S[0...n-1]$ each in the range $[u, v]$, it is not possible to construct a monotone bijective hash function ({\em minimal monotone perfect hash function}) that maps all the keys of the list into $j \in [0,n-1]$ without additional storage space less than $\Theta (n + \log w)$ bits~\cite{Belazzougui}. However, a bijective hash function can be constructed as a partial function~\cite{rosen:discrete_math_handbook} that assigns each key of $S_1 \subset S$ in the range $[\delta,\delta+n-1]$ with $\delta=\min(S)$ to exactly one element in $j \in [0,n-1]$. The partial hash function of this form is,
\begin{equation}\label{eqn:hash_func_algo}
\begin{split}
j=S[i]-\delta \quad \text{if} \quad S[i] - \delta < n
\end{split}
\end{equation}

With this definition, the proof has two basic steps of associative sort: 
\begin{enumerate}[label=\bf{Algorithm \Alph{*}.}, ref=Algorithm \Alph{*}, leftmargin=0pt, itemindent=*, start=1] 
\item \label{algorithm:es_fgl} Implicitly practice all the distinct keys in the range $[\delta, \delta+n-1]$ by mapping them into the imaginary subspace $Im[0...n-1]$ over $S[0...n-1]$ using Eqn.~\ref{eqn:hash_func_algo}. Assuming that the minimum of the list $\delta=\min(S)$ is known, this is,
\begin{enumerate}[label=\bf{A\arabic{*}.}, ref=A\arabic{*}, itemindent=*] 

\item  set $i = 0$;\label{algo2:item1}
\item if $S[i] - \delta \ge n$, then $S[i]$ is a key of $S_2$ that is out of the practiced interval. Hence, increase $n_d'$ that counts the number of keys in $S_2$, update $\delta'=min(\delta', S[i])$, increase $i$ and repeat this step;\label{algo2:item2}
\item if $i = S[i] - \delta$, then $S[i]$ is a node. Hence, increase $i$ and goto step \ref{algo2:item2};\label{algo2:item3}
\item otherwise, $S[i]$ is a key of $S_1$ to be practiced. Hence, swap $S[i]$ with $S[j]$ where $j = S[i] - \delta$. Increase $i$ if $j \le i$. Goto step \ref{algo2:item2};\label{algo2:item4}

\end{enumerate}

\item Implicitly store all the practiced keys which satisfy the monotone bijective hash function $i = S[i]-\delta$. If a key satisfies the hash function, then it is indeed  a node of the imaginary subspace. From algorithm point of view, this is partitioning the list into practiced and unpracticed keys, i.e.,
\begin{enumerate}[label=\bf{B\arabic{*}.}, ref=B\arabic{*}, itemindent=*]
\label{algorithm:fgp_distinct}
\item set $i = 0$, $j = 0$ and $k = n_d$;
\item if $i \ne S[i] - \delta$, then $S[i]$ is a key of $S_2$ that is out of the practiced interval. Increase $i$ and repeat this step;\label{algo3:item1}
\item otherwise, $S[i]$ is a node. Hence, swap $S[i]$ with $S[j]$. Increase $i$ and $j$ and decrease $k$. if $k = 0$ exit, otherwise goto step \ref{algo3:item1}.\label{algo3:item2}
\end{enumerate}

\end{enumerate}

As it is foreknown that all the keys of the list are distinct, assuming all $n_d$ keys in the practiced interval are located at wrong positions, $n_d$ keys are swapped while associations are created during implicitly practicing phase. Afterwards, those $n_d$ keys are stored (clustered) at the beginning of the list retaining their relative order, with at most $n_d-1$ swaps resulting in the sorted permutation of the practiced interval. On the other hand, if the keys in the range $[\delta, \delta+n-1]$ are consecutive, 
$$a,a+1, a+2,\dots$$
and already located at the beginning of the list in order, no any integer is swapped during practicing and  storing phases of associative sort.

\end{proof}

\lemref{lem:sorting_seq_of_distinct} proves that all the keys in a given interval spanned by the imaginary subspace can be sorted associatively in $\mathcal{O}(n)$ time using $\mathcal{O}(1)$ constant space, provided that the range of this interval is upper bounded by $n$. But the solution can be applied either sequentially or recursively to the entire list until all the keys are sorted. The sequential version is,
\begin{enumerate}[label=\bf{\roman{*}.}, ref=(\roman{*}), itemindent=*]
\item find $\min(S)$ and $\max(S)$;\label{algo4:item1}
\item initialize $\delta = \min(S)$, $\delta' = \max(S)$, $n_d'=0$;
\item implicitly practice all the distinct keys using \ref{algorithm:es_fgl};\label{algo4:item3}
\item implicitly store all the distinct keys using \ref{algorithm:fgp_distinct};\label{algo4:item4}
\item If $n_d = n$ exit. Otherwise, set $S=S[n_d-1 \ldots n-1]$, $n = n-n_d$, $\delta = \delta'$, $\delta' = \max(S)$, reset counters $n_d$ and $n_d'$ and goto step \ref{algo4:item3}.\label{algo4:item5}
\end{enumerate}

\begin{rem}
It should be noted that $\min(S)$ and $\max(S)$ need not be found in step \ref{algo4:item1}. Instead, if $\delta = 0$ and $\delta' = \max(\mathbb{U})$ the algorithm sorts the keys in the range $[0,n-1]$ during the first iteration (or recursion). However, if there is not any key in this interval, \ref{algorithm:es_fgl} finds $\delta'=\min(S)$ in step \ref{algo4:item3} in $\mathcal{O}(n)$ time, and the sorting continues with the keys in $[\delta', \delta'+n-1]$ in the next iteration (or recursion).
\end{rem}

\begin{rem}
Associative sort technique is on-line in the sense that after each step \ref{algo4:item4}, $n_d$ keys of $S_1$ are sorted at the beginning of the list and ready to be used.
\end{rem}

\begin{rem}
If there is satellite information along with the list $S$, they should be swapped together with the keys at step \ref{algo2:item4} and \ref{algo3:item2}. 
\end{rem}

\begin{description}[leftmargin = 0pt]

\item[{\bf Best Case Complexity.}] Given $n$ integer keys $S[0 \ldots n-1]$, if $n-1$ keys satisfy $S[i] - \delta < n$, then these keys are sorted in $\mathcal{O}(n)$ time. In the next step, there is one key left which implies sorting is finished. As a result, time complexity of the algorithm is lower bounded by $\Omega(n)$ in the best case.

\item [{\bf Worst Case Complexity.}] Given $n$ integer keys $S[0 \ldots n-1]$ and $m=\beta n$, if there is only $1$ key available in the practiced interval at each iteration (or recursion) until the last, in any $j$th step, the only key $s$ that will be sorted satisfies $s < jn-(j-1)$ which implies that the last alone key satisfies $s < jn-(j-1) \le \beta n$ from where we can calculate $j$ by $j \le \frac{\beta n-1 }{n-1}$.
In this case, the time complexity of the algorithm is,
\begin{equation}
\mathcal{O}(n) + \mathcal{O}(n-1) + \dotsc + \mathcal{O}(n-j) = (j+1) \mathcal{O}(n) -\mathcal{O}(j^2) < (\beta+1) \mathcal{O}(n)
\end{equation}

Therefore, the algorithm is upper bonded by $(\beta+1) \mathcal{O}(n) = \mathcal{O}(m+n)$ in worst case.

\item[{\bf Average Case Complexity.}] Given $n$ integer keys $S[0 \ldots n-1]$, if $m = \beta n$ and the keys are uniformly distributed, this means that $\frac{n}{\beta}$ keys satisfy $S[i] < n$. Therefore, the algorithm is capable of sorting $ \frac{n}{\beta}$ keys in $\mathcal{O}(n)$ time during first pass. This will continue until all the keys are sorted. The sum of sorted keys in each iteration can be represented with the series,
\begin{equation} \label{eqn:series_sorted_case3}
\frac{n}{\beta}+\frac{n(\beta-1)}{\beta^2}+\dotsc+\frac{n(\beta-1)^{k-1}}{\beta^{k}}+\dotsc
\end{equation}

It is reasonable to think that the sorting ends when one term is left which means the sum of $k$ terms of this series is equal to $n-1$, from where we can calculate the number of iteration or dept of recursion $k$ which is valid when $\beta > 1$ by,
\begin{equation} \label{eqn:series_sorted_case3_4}
\frac{1}{n} = \frac{(\beta-1)^{k-1}}{\beta^{k}}
\end{equation}
It is seen from Eqn.~\ref{eqn:series_sorted_case3_4} that when $m = 2n$, i.e., $\beta=2$, number of iteration or dept of recursion becomes $k=\log{n}$ and the complexity is the recursion $T(n) = T(\frac{n}{2}) + \mathcal{O}(n)$ yielding $T(n) = \mathcal{O}(n)$. It is known that each step takes $\mathcal{O}(n)$ time. Therefore, the time complexity of the algorithm is,
\begin{equation}\label{eqn:series_complexity_case3}
\begin{split}
\mathcal{O}(n)+\mathcal{O}\bigl(\frac{n(\beta-1)}{\beta}\bigr) +\dotsc +\mathcal{O}\bigl( \frac{n(\beta-1)^{k-1}}{\beta^{k-1}} \bigr)
\end{split}
\end{equation}
from where we can obtain by defining $x= \frac{(\beta-1)}{\beta}$,
\begin{equation}\label{eqn:ud_6}
\mathcal{O}(n) \bigl( 1 +  x + x^2 + x^3 + \cdots + x^{k-1} \bigr) = \mathcal{O}(n) (\frac{1}{1-x} - \frac{x^{k-1}}{1-x}) < \beta \mathcal{O}(n)
\end{equation}
which means that the algorithm is upper bounded by $ \beta \mathcal{O}(n)$ or $\mathcal{O}(m)$ in the average case.

\item[{\bf More on Complexity}] The time complexity of the overall algorithm is upper bounded by $\mathcal{O}(n+m)$ and lower bounded by $\Omega(n)$. Upper bound $\mathcal{O}(n+m)$ does not mean that the complexity is unbounded when $m > n^c$ with $c\ge 2$. In each iteration, $\min(S_2)$ is found and the new iteration starts with this minimum. Hence, the number of iteration or recursion is bounded by the number of keys, i.e., $k \le n$. This means that the complexity is upper bounded by $\mathcal{O}(n^2)$ regardless of how much big is the ratio $\frac{m}{n}$.

Number of total swaps is at most $2n-k$ because in each implicitly practicing and implicitly storing phases at most $2n_d-1$ keys of $S_1$ are swapped which sum to $2n-k$ overall where $k\ge1$ is the number of iteration (or dept of recursion) to complete the sorting.  If $k=n$, then the the list is sorted with at most $n-1$ swaps in $\mathcal{O}(n^2)$ time.

If the list is already sorted and the keys of $S_1$ are always consecutive or only one key is available in the practiced interval all the time, then no any swap occurs.

\end{description}


\section{Sorting $n$ Distinct Modifiable Integers}\label{subsec:es_multiple}

In this section, the associative sorting technique for distinct modifiable integers will be introduced with its three basic steps: (i) practicing, (ii) storing and (iii) retrieval. This is an integer sorting problem not an integer key sorting problem. Hence, the definition degenerates to: given a {\em list} $S$ of $n$ {\em integers}, $S[0 \ldots n-1]$, the problem is to sort the integers in ascending or descending order.

Once a node is created for a particular integer when it is practiced, the redundancy due to the association between the integer and the node releases the word allocated for the integer in the physical memory except one bit which is used to tag the word as node of the imaginary subspace for interrogation. The released $w-1$ bits of a node become its record. Hence, we can improve the associative sort for {\em distinct} integers using the record of each node for other purposes. After practicing, each node is clustered at the beginning of the list together with its record retaining its relative order with respect to others during storing. At this point, we need $\log n$ bits of the record to encode the node's absolute position as the cue. But the tag bit is released during storing since we only need to know how many nodes are  stored at the beginning of the list in total. Hence we can use $w-\log n$ bits of a record during practicing for any other purpose. It is immediate from this definition that,
\begin{lem}\label{lem:sorting_seq_of_distinct_m}
Given $n$ distinct integers $S[0...n-1]$ each in the range $[u, v]$, all the $n_d$ integers in the range $[\delta,\delta+(w-\log n)n-1]$ with $\delta=\min(S)$ can be sorted associatively at the beginning of the list in $\mathcal{O}(n)$ time using only $\mathcal{O}(1)$ constant space.
\end{lem}

Given $n$ distinct integers $S[0...n-1]$ each in the range $[u, v]$, a monotone bijective {\em super} hash function can be constructed as a partial function that assigns each integer of $S_1 \subset S$ in the range $[\delta,\delta+(w-\log n)n-1]$ with $\delta=\min(S)$ to exactly one element in $j \in [0,n-1]$ and one element in $k \in [0,(w-\log n)-1]$. The simplest monotone bijective partial super hash function of this form is,
\begin{equation}\label{eqn:shf_1}
j = (S[i] - \delta) \, \idiv \, (w - \log n)  \quad  \text{if} \quad S[i] - \delta < (w-\log n)n  
\end{equation}
\begin{equation}\label{eqn:shf_2}
k = (S[i] - \delta) \imod (w - \log n)   \quad \text{if} \quad S[i] - \delta < (w-\log n)n 
\end{equation}
In this case, $w - \log n$ integers may collide and mapped to the same node created at $j \in [0,n-1]$ (Eqn.~\ref{eqn:shf_1}) in the imaginary subspace. But we can use $w- \log n$ free bits of a record of the node to encode which of $w-\log n$ distinct integers are mapped to the same node by setting the corresponding bit determined by $k$ (Eqn.~\ref{eqn:shf_2}). In other words, now the imaginary subspace is two dimensional over the list space where the first dimension along the list designates the node position and the second dimension along the bits of the node uniquely determines the integers which are mapped to the imaginary subspace through that node.

\begin{proof}
With this definition, the proof has three basic steps of associative sort:

\begin{enumerate}[label=\bf{Algorithm \Alph{*}.}, ref=Algorithm \Alph{*}, leftmargin=0pt, itemindent=*, start=3] 
\item \label{algorithm:es_fgl_m} Practice all the distinct integers of the interval $[\delta,\delta+(w-\log n)n-1]$ by mapping them to the node determined by Eqn.~\ref{eqn:shf_1} in the imaginary subspace $Im[0...n-1]$ over $S[0...n-1]$. Once a integer is mapped to a node, set the integer's unique bit in the record determined by Eqn.~\ref{eqn:shf_2} which discriminates it from the others mapped to the same node. It is assumed that minimum of the list $\delta = \min{S}$ is known.
\end{enumerate}

\begin{enumerate}[label=\bf{C\arabic{*}.}, ref=C\arabic{*}, itemindent=*]
\item set $i = 0$;\label{algo5:item0}
\item if $S[i] < \delta$, then $S[i]$ is an idle integer of an interval that has already been sorted in the previous iterations (or recursions). Hence, increase $i$ and repeat this step;\label{algo5:item1}
\item if MSB of $S[i]$ is $1$, then $S[i]$ is a node. Hence, increase $i$ and goto step \ref{algo5:item1};\label{algo5:item2}
\item if $S[i] - \delta \ge (w-\log n)n$ then $S[i]$ is a integer of $S_2$ that is out of the practiced interval. Increase $n_d'$ that counts the number of integers of $S_2$, update $\delta'=min(\delta', S[i])$, increase $i$ and goto to step \ref{algo5:item1};\label{algo5:item3}
\item calculate $j$ and $k$ using Eqn.~\ref{eqn:shf_1} and \ref{eqn:shf_2}, respectively;\label{algo5:item4}
\item if MSB of $S[j]$ is $0$, then $S[i]$ is the first occurrence which will create the node at $S[j]$. Hence, move $S[j]$ to $S[i]$, clear $S[j]$ and set MSB and $k$th bit of $S[j]$ to $1$. If $j \le i$ increase $i$. Increase $n_d$ that counts the number of distinct integers and hence the nodes, and goto step \ref{algo5:item1}.\label{algo5:item5}
\item otherwise, a node has already been created at $S[j]$ by another occurrence of $S[i]$. Hence, set $k$th bit of $S[j]$ (without touching others) and increase $i$ and $n_c$ that counts number of total idle integers over all distinct integers, and goto step \ref{algo5:item1};
\end{enumerate}

\begin{enumerate}[label=\bf{Algorithm \Alph{*}.}, ref=Algorithm \Alph{*}, leftmargin=0pt, itemindent=*, start=4] 
\item \label{algorithm:es_fgp_m} Store all the integers of the practiced interval in the short term memory by clustering the nodes at the beginning of the list while retaining their relative order with respect to each other. During clustering encode absolute position of each node into its record's most significant $\log n$ bits.
\end{enumerate}
\begin{enumerate}[label=\bf{D\arabic{*}.}, ref=D\arabic{*}, itemindent=*]
\item set $i = 0$, $j = 0$ and $k = n_d$;
\item if MSB of $S[i]$ is $0$, then $S[i]$ is not a node, hence increase $i$ and repeat this step;\label{algo6:item1}
\item if MSB of $S[i]$ is $1$, then $S[i]$ is a node. Hence, clear MSB of $S[i]$, encode $i$ into most significant $\log n$ bits of $S[i]$ and swap $S[i]$ with $S[j]$. Increase $i$ and $j$ and decrease $k$. If $k = 0$ exit, otherwise goto step \ref{algo6:item1}. \label{algo6:item2}
\end{enumerate}

\begin{enumerate}[label=\bf{Algorithm \Alph{*}.}, ref=Algorithm \Alph{*}, leftmargin=0pt, itemindent=*, start=5] 
\item \label{algorithm:es_fgd_m} Retrieve $n_d+n_c$ integers of $S_1$ from $n_d$ records of short term memory $S[0 \ldots n_d-1]$ to construct sorted permutation of $n_d+n_c$ integers of $S_1$. Process the records from right to left backwards and expand the integers over $S[0 \ldots n_d+n_c-1]$ sequentially right to left backwards.
\end{enumerate}
\begin{enumerate}[label=\bf{E\arabic{*}.}, ref=E\arabic{*}, itemindent=*]
\item initialize $i = n_d-1$ and $p = n_d + n_c -1$;
\item initialize $k = w-\log n-1$ and decode absolute position $j$ of the node from most significant $\log n$ bits of its record at $S[i]$;\label{algo7:item1}
\item while $k \ge 0$; 
\begin{enumerate}[label=(\roman{*})]
\item if $k$th bit of $S[i]$ is $1$, then $S[p] = j(w - \log n) + k + \delta$. Decrease $k$ and $p$ and repeat this step. 
\item otherwise, only decrease $k$ and goto step (i).
\end{enumerate}
\item decrease $i$ and goto step \ref{algo7:item1};\label{algo7:item2}
\end{enumerate}

\end{proof}

\begin{description}[leftmargin = 0pt]

\item [{\bf Sequential Version}] Instead of using stack space, a sequential version can be developed. After storing, $n_c$ idle integers of $S_1$ and $n_d'$ integers of $S_2$ are distributed disorderly together at $S[n_d \ldots n-1]$. If we partition $S[n_d \ldots n-1]$ selecting the pivot equal to $\delta$, then idle integers are clustered after $n_d$ records of the short term memory. Therefore, \ref{algorithm:es_fgd_m} can immediately be used to retrieve. In such a case, the structure of the overall algorithm becomes:
\begin{enumerate}[label=\bf{\roman{*}.}, ref=(\roman{*}), itemindent=*]
\item find $\min(S)$ and $\max(S)$;\label{algo8:item1}
\item initialize $\delta = \min(S)$, $\delta' = \max(S)$, $n_d=0$, $n_c=0$, $n_d'=0$;\label{algo16:item2}
\item practice all the integers using \ref{algorithm:es_fgl_m};\label{algo16:item3}
\item store all the practiced integers using \ref{algorithm:es_fgp_m};\label{algo16:item4}
\item partition $S[n_d\ldots n-1]$ clustering $n_c$ idle integers of $S_1$ to the beginning using any in-place partitioning algorithm;\label{algo16:item5}
\item retrieve the sorted permutation of the practiced interval using \ref{algorithm:es_fgd_m};\label{algo16:item6}
\item If $n_d'=0$ exit. Otherwise set $S=S[n_d+n_c \ldots n-1]$. Update $n = n-n_d$, $\delta = \delta'$, $\delta' = \max(S)$, reset counters $n_d$, $n_c$ and $n_d'$ and goto step \ref{algo16:item3}.\label{algo16:item7}
\end{enumerate}

\begin{rem}
$\min(S)$ and $\max(S)$ need not be found in step \ref{algo8:item1}. Instead, if $\delta = 0$ and $\delta' = \max(\mathbb{U})$ the algorithm sorts the integers in the range $[\delta,\delta+(w-\log n)n-1]$ during the first iteration (or recursion). However, if there is not any integer in this interval, \ref{algorithm:es_fgl_m} finds $\delta'=\min(S)$ in step \ref{algo5:item3} in $\mathcal{O}(n)$ time, and continues with the integers in $[\delta',\delta'+(w-\log n)n-1]$.
\end{rem}

\begin{rem}
Sequential version of associative sort technique is on-line in the sense that after each step \ref{algo16:item6}, $n_d+n_c$ integers are added to the sorted permutation at the beginning of the list and ready to be used.
\end{rem}

\item [{\bf Recursive Version}] Saving constants $n$, $n_d$, $n_c$ and $\delta$ in stack space, we can recursively call \ref{algorithm:es_fgl_m} and \ref{algorithm:es_fgp_m} with three parameters $n = n_c + n_d'$, $S=S[n_d \ldots n-1]$ and $\min(S) = \delta'$. The last parameter prevents searching the minimum of the list in each level of recursion. Although the exact number of integers of $S_2$ is $n_d'$, the number of integers of $S$ in the new recursion is $n_c+n_d'$ where $n_c$ of them are idle integers of $S_1$ and meaningless while sorting $S_2$. However, this will increase the interval of range of integers spanned by the imaginary subspace improving the overall time complexity in each level of recursion. The recursion can continue until no any integer exists in $S_2$. In the last recursion, retrieval phase can begin to construct the sorted permutation of $n_d+n_c$ integers of $S_1$ from $n_d$ records stored at the short term memory $S[0 \ldots n_d-1]$ and expand over $S[0 \ldots n-1]$ sequentially right to left backwards. Each level of recursion should return the total number of integers expanded on the list to the higher level to let it know where it will start to expand its interval.   

\end{description}

\begin{description}[leftmargin = 0pt] 
\item[{\bf Worst Case Complexity}] \ref{algorithm:es_fgl_m}, \ref{algorithm:es_fgp_m} and \ref{algorithm:es_fgd_m} are together capable of sorting integers that satisfy $S[i] - \delta < (w-\log n)n$ in $\mathcal{O}(n)$ time. If we assume $m = \beta n$ with $\beta > w-\log n$, and there is only one integer available that satisfies $S[i] - \delta < (w-\log n)n$ in each iteration or recursion until the last, in any $j$th step, the only integer $s$ of $S_1$ that will be sorted satisfies,
\begin{equation}\label{eqn:distinct_m_wc}
s - \delta < j(w-\log n)n-(j-1)
\end{equation}
Eqn.~\ref{eqn:distinct_m_wc} implies that the last alone integer of $S$ satisfies,
\begin{equation}\label{eqn:distinct_m_wc_1}
s - \delta  < j(w-\log n)n-(j-1) \le \beta n
\end{equation}
from where can calculate $j$ by,
\begin{equation}
j \le \frac{\beta n-1}{(w-\log n)n - 1}
\end{equation}
In this case, the time complexity of the algorithm is 
\begin{equation}\label{eqn:distinct_m_wc_2}
\begin{split}
\mathcal{O}(n) + & \mathcal{O}(n-1) + \dotsc  + \mathcal{O}(n-j) = (j+1) \mathcal{O}(n) -\mathcal{O}(j^2) < (\frac{\beta}{w-\log n}+1) \mathcal{O}(n)
\end{split}
\end{equation}
Therefore, the time complexity of the algorithm in worst case is upper bounded by $\mathcal{O}(n+\frac{m}{w-\log n})$.


\item[{\bf Best Case}] If $n-1$ integers satisfy $S[i] - \delta < (w-\log n)n$, then these are sorted in $\mathcal{O}(n)$ time. In the next step, there is $n'=1$ integer left which implies sorting is finished. As a result, time complexity of the algorithm is lower bounded by $\Omega(n)$ in the best case.


\item[{\bf Average Case}] If we assume $m = \beta n (w-\log n)$ with $\beta> 1$, and the integers are uniformly distributed, this implies $\frac{n}{\beta}$ integers satisfy $S[i] - \delta < (w-\log n)n$. Therefore, the algorithm is capable of sorting $ \frac{n}{\beta}$ integers of the list in $\mathcal{O}(n)$ time at first step and $n'=n-\frac{n}{\beta}=\frac{n(\beta-1)}{\beta}$ integers will be left where $\frac{n'}{\beta}$ of them will be sorted in the next step. This will continue until all the integers are sorted. The complexity in this case is exactly equal to the complexity that we obtained for associative sorting of read-only distinct integers, which means that the time complexity of the sorting algorithm is upper bounded by $ \beta \mathcal{O}(n)$ or $\mathcal{O}(\frac{m}{w-\log n})$ for uniformly distributed lists.

\item[{\bf Practical Experience}] When modifiable version is compared with read-only version, it has been observed that up to $\frac{m}{n} \le 10$ read-only version is more efficient possibly due to bitwise operations involved in the modifiable version. Afterwards modifiable version becomes more efficient than read-only version augmenting it.

\end{description}

\section{Conclusions}
\label{chap:summaryandconclusion}

In this study, in-place associative integer sorting technique is specialized with two variants one for read-only integer keys and the other for modifiable integers. Both techniques are very simple and straightforward and around 30 lines of C code is enough.

The read-only variant sorts the keys using $\mathcal{O}(1)$ extra space in $\mathcal{O}(n+m)$ time for the worst, $\mathcal{O}(m)$ time for the average (uniformly distributed keys) and $\mathcal{O}(n)$ time for the best case. It shows similar characteristics with bucket sort and distribution counting sort but it is in-place and hence time-space efficient. The ratio $\frac{m}{n}$ defines the efficiency (time-space trade-offs) letting very large lists to be sorted in-place. Hence, since it does not require additional memory space other than a constant amount, no matter how large is the list, the proposed algorithm will sort faster than {\em all} provided that $m = \mathcal{O} (n \log n)$. It should be noted that, this variant is suitable for integer key sorting problems where satellite information is available along with the integer keys.

On the other hand, the modifiable integer variant sorts the integers using $\mathcal{O}(1)$ extra space in $\mathcal{O}(n+\frac{m}{w-\log n})$ time for the worst,  $\mathcal{O}(\frac{m}{w-\log n})$ time for the average (uniformly distributed keys) and $\mathcal{O}(n)$ time for the best case. It should be noted that, this variant is not suitable for integer key sorting problems where satellite information is available along with the integer keys. It is suitable for integer sorting problems. It has been observed practically that up to $\frac{m}{n} \le 10$ read-only version is more efficient possibly due to bitwise operations involved in the modifiable version. Afterwards modifiable version becomes more efficient than read-only version augmenting it.

\end{document}